# Catch the Patient, Not the AI: Collective Sensemaking in an Online Health Community

Catch the Patient, Not the AI

A mixed-methods study of a Chinese lymphoma patient forum

Feng He*

Independent researcher, hefeng.research@gmail.com

Patients and caregivers increasingly use artificial intelligence (AI) tools to interpret medical reports, weigh care decisions, and seek emotional support. Yet most research treats patient-facing AI as a private exchange between a user and a system, leaving open how AI-related content is taken up once users carry it back into the peer communities where they already make sense of illness. This study examines collective sensemaking around AI in House086, China's largest online community for lymphoma patients and caregivers. We identified roughly 400 publicly accessible discussion threads (2014 to 2026) through AI-related keyword searches and manual screening, extracted them into structured case profiles using a schema-prompted large language model (Claude Sonnet 4.6, deployed via AWS Bedrock), and conducted mixed-method analysis. After quality control, the verified analytic sample comprised 337 post-ChatGPT records focused on widespread conversational-AI use. Members most often reported using AI for informational support, followed by second opinions and psychosocial support. Although members often introduced AI favorably, roughly one in six described feeling overwhelmed by AI output. When other members responded, they frequently engaged the poster's underlying medical or emotional intent while leaving the AI dimension unaddressed. The tendency to bypass AI persisted even in threads seeking triangulation between AI output and other information sources, and when posters shared overwhelmed or unfavorable experiences with AI. On the occasions when members did discuss the AI, they were more often cautious than endorsing. Rather than auditing AI output for factual accuracy, the community more often worked to deflate the false certainty it produced, placing a single AI answer back among multiple sources of judgment. This study argues that AI does not replace the interpretive work of online health communities, nor is it systematically audited by them. Instead, it shifts the locus of sensemaking downstream, so that the community continues to catch the person even when it does not catch the AI, and the interpretive authority that follows does not settle in any single place. We discuss implications for CSCW theory, online-community norms, and patient-facing AI design.

**Additional Keywords and Phrases:** Online health communities, Collective sensemaking, Patient-facing AI, Large language models, Peer support, Computer-supported cooperative work

---

## 1 INTRODUCTION

"Emergency: can someone help read these numbers?" A frightened patient had posted to an online lymphoma forum, overwhelmed about their condition. Part of their anxiety came from an AI chatbot, which had analyzed their medical reports and returned an alarming interpretation. Community members quickly responded by checking the uploaded screenshots of the same reports, reassuring the patient, and offering practical suggestions grounded in the patient's condition. Among these replies, however, only one briefly suggested that the AI may have misread the report. The thread then moved on.

Online health communities (OHCs) have long served as a critical infrastructure for patients, caregivers, and the wider public to make sense of illness outside formal clinical encounters [1-4]. Patients and caregivers rely on them to navigate complex, high-stakes diseases such as cancer, where diagnosis and treatment carry substantial burden and uncertainty. In doing so, OHCs do more than circulate information or provide reassurance. They become a place where fragmented, technical, and personally consequential health information is interpreted collectively, by people who are living through the same disease.

Artificial intelligence (AI) tools are now entering this established ecology of peer support and illness interpretation. A growing share of the public reports turning to AI chatbots for health information and advice, raising concerns and discussions regarding its risks and potentials [5]. Recent HCI work has also begun to examine how patients incorporate LLMs into healthcare-seeking journeys [6]. To date, however, AI in health has largely been studied either as a standalone decision support system, or as an individual interaction between a user and a tool [6-10]. Far less is known about what happens next, after AI-related content becomes part of peer interaction, when users bring it back into the communities where they already interpret illness together.

This gap matters because patient-facing AI does not operate in isolation. In everyday illness work, a user might consult AI while seeking formal care, rely on AI to mediate social interactions, or recommend AI outputs for others to weigh in on. The community context shapes what such content comes to mean. Where AI claims are routinely promoted but rarely questioned, an individual may overestimate their credibility. Where claims are carefully examined and corrected, that member may benefit from collective interpretive labor that no single user could perform alone. Understanding these dynamics is necessary if we are to study AI as part of a wider social infrastructure of health-information seeking.

This study examines collective sensemaking around AI in House086, China's largest online community for lymphoma patients and caregivers. Founded in 2011, the forum now has over 166,000 members and more than three million threads [11]. This setting is analytically useful for two reasons. First, lymphoma is a blood cancer marked by diagnostic uncertainty but also by comparatively established treatment pathways. New patients and caregivers often arrive with questions about diagnosis, treatment choices, and what to do next, while many who have moved beyond the initial crisis continue to return to the forum, carrying forward experience that remains useful to others. As a result, the forum sustains large volumes of daily traffic and a wide range of discussions about living with illness. Second, in China's health system, fragmented care pathways and limited clinical resources often make patient activation practically necessary [6, 12]. As AI tools such as DeepSeek and Doubao become increasingly available to ordinary users, House086 becomes one of the few places where healthcare context, lived experience, and now AI-related content can meet. This makes the forum a useful site for examining how AI enters an existing online health community and becomes part of ordinary peer discussion.

In this study, approximately 400 publicly accessible House086 threads involving AI-related discussion were identified and analyzed through quantitative description and close reading of individual cases. The analysis was organized around three research questions. First, how did AI enter the forum, including who introduced it and for what purpose? Second, when a member brought AI into a thread, how did the community interact with that member? And third, how did the community approach AI as a topic?

The study makes three contributions. First, it provides empirical evidence of how AI enters everyday discussion in a large disease-specific OHC, often as part of ordinary illness sensemaking rather than as a stand-alone technology object. Moreover, it offers an analytic distinction between two dimensions of community response: how members respond to the person who brings AI into discussion, and how they approach AI as a topic. Finally, it shows that these two dimensions can diverge. The community often prioritizes the situated illness work of the person seeking help, while leaving AI outside the center of collective attention or norm negotiation. Taken together, these results contribute to CSCW and HCI research on OHCs by showing how an established patient community reliably catches the patient, not the AI.

The remainder of the paper proceeds as follows. Section 2 reviews related work on collective sensemaking in OHCs, patient-facing AI for individual users, and how AI began to enter social platforms. Section 3 describes our methods. Section 4 presents our findings, organized by the three research questions, with each part combining quantitative patterns and qualitative case reading. Section 5 discusses implications for OHC norms, CSCW, and patient-facing AI design. Section 6 concludes.

## 2 RELATED WORK

### 2.1 Online Health Communities as Sites of Collective Sensemaking

Collective sensemaking examines how groups make meaning together when facing ambiguous, dynamic, or uncertain situations [4]. This lens matters because some objects only become interpretable when people collectively work on them. Online health communities have long been studied as social spaces for collective sensemaking around illness, where patients, caregivers, and other stakeholders seek information, exchange experiential knowledge, and provide support outside formal medical care. Mamykina et al., in their study of TuDiabetes, showed how members of an online diabetes forum engaged in collective sensemaking through deep discussion, back-and-forth negotiation of perspectives, and resolution of conflicts in opinions [4]. They identified key attributes of the process, including lateral engagement, reflection on previous perspectives, and transformation of ideas. Moreover, they found that health forum participants often valued multiplicity and diversity of perspectives instead of a single agreed-upon answer.

A second body of work clarifies why people come to these communities in the first place. Studies of participation in OHCs and social support online distinguish a recurring set of motivations: seeking information, seeking emotional or esteem support, sharing one's own experiential knowledge, and lighter forms of companionship and sociability that sustain the community over time [13, 14]. Wang, Kraut, and colleagues, for instance, show that informational and emotional support play different roles in why members join, contribute, and stay [14].

A related line of research examines the distinctive expertise that patients and peers contribute. Hartzler and Pratt argue that patient expertise differs from clinician expertise because it is grounded in lived experience, day-to-day management, and practical adaptation to illness [1]. This form of expertise allows community members to interpret medical information in relation to everyday circumstances, compare experiences across similar trajectories, and address concerns that may fall outside the scope of medical authority. Huh and colleagues similarly show that OHCs can support the "whole person," weaving together medical, emotional, social, and practical concerns that do not map neatly onto clinical categories [2]. Further studies of how moderation and clinical expertise are woven into patient communities highlight the ongoing negotiation of boundaries between peer knowledge and professional authority [3].

This literature is important for the present study because AI-related content is entering OHCs. Existing research provides the foundation for understanding community norms and practices for evaluating information, sharing experience, and managing uncertainty, but it largely predates the everyday use of consumer AI tools by patients and caregivers. In OHCs, members do not encounter new information sources in a vacuum. It therefore remains an open question on how AI-related content becomes part of collective sensemaking in peer discussion.

### 2.2 Patient-Facing AI for Health-Information Use

Large language models (LLMs) are increasingly used as patient-facing tools for interpreting health information. A national tracking poll reported that many adults have used AI chatbots for health information or advice, some uploading personal materials such as test results or clinicians' notes [5]. A growing body of research has begun to evaluate such use. In the OpenNotes context, Salmi and colleagues assessed LLM responses to patient questions about clinical visit notes and found that such models may help patients understand their health information, while stressing the need for careful guidance in their use [15]. Extending beyond single episodes of information interpretation, Cao et al. conducted a four-week diary study with Chinese patients, showing that LLMs can take on behavioral, informational, emotional, and cognitive roles across healthcare seeking trajectories [6].

A second strand moves from patient agency to the quality, appeal, and risks of AI responses to patient questions. Benchmarking studies report that LLM-generated answers are often rated as more empathetic than clinician responses [7, 16]. This perceived warmth may be part of why people turn to AI for reassurance, yet the same fluency can mislead. A separate literature has documented safety concerns in which confident, well-phrased AI output was incorrect in ways that could plausibly lead to harm if acted on without oversight [8], underscoring that empathy and surface quality are not proxies for reliability. However, other work shows that LLMs may serve a protective function, for example by flagging possible diagnostic errors or inconsistencies that a patient or even a clinician might otherwise miss [9].

These studies together point to an open question about where the corrective work happens. If AI can catch some errors but also introduce others, the question of who notices, contextualizes, and acts on a given AI output becomes consequential, and that question is social rather than purely technical. Much of this literature, however, still treats AI use as an individual interaction between a user and a system, asking whether its responses can support a patient or clinician acting alone [6-9].

### 2.3 AI, Support, and Norms in Online Communities

As AI-generated writing becomes harder to distinguish from human participation, online communities face new concerns about transparency, credibility, and trust. Work on authorship perception shows that participants may struggle to tell whether a health-related response was written by a human or an AI, complicating community norms around lived experience, accountability, and peer support [17]. Such uncertainty is especially consequential in health communities, where advice and reassurance can shape how members interpret symptoms, treatment options, and professional guidance.

Beyond authorship and disclosure, a strand of work asks whether AI-generated responses can provide appropriate support in online health communities. Mittal and colleagues show that AI-generated support cannot be assessed by surface helpfulness or linguistic quality alone, but must also be evaluated against the norms and expectations of the specific community in which it is offered [18]. This frames AI support as a community-embedded phenomenon rather than only a user-system interaction. Related studies of conversational agents and AI-enhanced support infrastructures similarly raise questions about trust, empathy, moderation, and the boundary between peer and automated support [19].

Most existing work in this area treats AI as a visible participant in peer discussion, or as a platform infrastructure providing direct support. The resulting analytic focus is the community's collective encounter with AI. In everyday practice, however, AI-related content may enter an OHC by a less visible route. Individual members may first use AI privately outside the community, before bringing AI-generated outputs, emotional reactions, and provisional judgments back for discussion. Here AI is no longer an independent participant, but is present through the member whose narrative has been shaped by it. This carried-in form of AI is structurally left out by existing research designs, where the unit of analysis is the AI-authored contribution itself, and the research question is how the community evaluates it. When AI instead enters the discussion through a member's narrative, there is no AI-authored unit to code. What then requires analysis is the member-thread interaction that carries it. The present study therefore takes the member who brings AI into a thread, together with the surrounding thread-level interaction, rather than any AI-authored post, as its unit of analysis.

# 3 METHODS

## 3.1 Study Sites and Thread Identification

This study used publicly accessible threads from house086.com ("Lymphoma Home"), the largest online community for lymphoma patients and caregivers in China [11]. Founded in 2011, the forum hosts over three million discussion threads on lymphoma diagnosis, treatment, and lived experience, with more than 166,000 registered members at the time of data collection in May 2026. Members include patients, family caregivers, forum administrators and, on occasion, clinicians posting through official channels.

Candidate threads were identified through systematic keyword searches of the forum's full-text search interface. Pilot reading suggested that forum users used broad and inconsistent labels for AI-related tools. Restricting the search to specific generative AI models would therefore have imposed an analyst-defined taxonomy that the community itself does not use, and would have missed posts that are about AI in the members' own sense of the word. The final search terms accordingly combined generic references to AI with names of specific tools. Generic terms covered the common Chinese and English ways members referred to AI technologies, large models, and machine learning. Tool-specific terms covered conversational AI assistants active in the Chinese market, such as DeepSeek, ChatGPT, Doubao, Kimi, and Qwen, among others, together with several medical-oriented assistants. After the forum’s own internal AI chatbot was identified as an OHC-deployed AI artifact through manual reading, it was also added as a search target. The full search-term list appears in **Appendix A.1.1**.

Further, this study avoided restricting the search by date, so that the contemporary AI-related threads could be interpreted against earlier discussions of AI in the forum. In practice, substantive AI-relevant threads were posted after 2022 and concerned contemporary conversational AI tools.

## 3.2 Coding Schema and LLM-Assisted Case Extraction

Following initial human screening, each retained thread was coded against a structured schema using LLMs. The schema, developed through manual reading and iterative refinement, models each thread as an interaction among three parts: the initiator, defined as the first OHC member who explicitly brings AI into a thread; the AI-related content, defined as a tool, output, or interpretation being introduced; and the community, defined as the set of members who respond, or do not respond, to the initiator's post. In the coding schema, the initiator and AI-related fields characterize what enters the OHC, while the community fields characterize how the group responds.

For the initiator, extracted variables include their role in the community (patient, caregiver, or community authority), their shared experience with AI, and their intent in posting, each accompanied by supporting evidence quoted from the original text. Authority-coded initiators included forum administrators, official editorial accounts, and clinician voices behind them; the forum's internal AI bot was also coded here, since it is platform infrastructure controlled directly by administrators. Experience was coded as favorable, neutral, overwhelmed, or unfavorable as reflected in their post. The "overwhelmed" category, developed after initial manual case reading, covers posts in which AI output triggered or intensified emotional collapse or information overload, paralyzing the initiator's capacity to interpret or decide. This differed from unfavorable experiences, which referred to posts that rejected or criticized AI as misleading, harmful, or untrustworthy. Posting intents were coded as seeking information, seeking psychosocial support, seeking triangulation of conflicting advice, sharing information, and sharing experience or sociable play, according to the established accounts of OHC participation and social support [13, 14].

For the community, the response was characterized along two dimensions. The first dimension asks whether members took up the initiator's posting intent (yes, partial, or no), drawing on the idea of lateral engagement [4]. The second dimension captures the community's stance toward the AI-related content (supportive, cautious, dismissive, mixed, or not addressed). This stance dimension was developed inductively during manual reading, with "not addressed" defined as cases where members did not comment on the AI dimension at all.

For the AI-related content, the tool or tools referenced and the functional use to which AI was put were recorded. Functional uses were coded as non-exclusive categories including information support, second opinion, psychosocial support, and other. Information support referred to using AI to explain, search for, organize, or navigate medical and care-related information. Second opinion referred to using AI to check, challenge, or support recommendations or decisions from clinicians or hospitals. Psychosocial support included uses of AI for comfort, companionship, encouragement, emotional expression, or playful and creative activities. The residual "other" category was used when none of these three categories applied, and all cases coded as "other" were subsequently reviewed manually to ensure coding consistency.

Because a single thread can reference multiple AI tools and draw heterogeneous responses, the unit of extraction is one initiator case per thread. For each thread, the study additionally recorded the time of posting and a five-point case-richness score indicating the depth of peer interaction. Operational definitions for all fields appear in **Appendix A.1.2.**

Claude Sonnet 4.6 was chosen to assist with automatic extraction of structured records for its capacity to handle long, complex conversational data and structured JSON output. Other frontier LLMs, such as GPT and DeepSeek, were avoided because they frequently appeared as objects of discussion in these threads, and using them to infer initiator experiences and community stances toward themselves may introduce a potential source of bias. Deployed on AWS Bedrock, Claude (temperature=0.1) received a detailed coding prompt containing operational definitions, anchor examples, and an explicit instruction to process raw thread HTML and OCR text recovered from embedded screenshots into a single structured JSON record per thread, or null where no substantive AI mention was present. Each thread was processed as an independent request to avoid context contamination and ensure extraction consistency. All model outputs and extraction failures were reviewed by the author against the original thread content with failed or incomplete cases repaired manually where needed. Independent double coding was not conducted at this stage and is treated as a limitation.

### 3.3 Sample Composition and Analysis Approach

Initial human screening removed inaccessible, duplicate, and clearly irrelevant threads, retaining 409 candidate threads for LLM-assisted extraction and manual verification. Of these, 384 threads were successfully converted to JSON records, six outputs with parsing errors were repaired using a JSON-repair fallback, and six threads were manually patched. The remaining 13 threads with trivial or irrelevant AI mentions, such as uses of "AI" to refer to Adobe Illustrator, were excluded (**Table 1**).

Among the 396 verified records with substantial AI-related content, 45 records originating from a single highly active promotional-style account were excluded from substantive analysis but retained in the descriptive account of the forum's data ecology (Section 4.1). The remaining 351 records, spanning March 2014 to May 2026, included 14 records posted before the public release of ChatGPT in November 2022. As these pre-ChatGPT threads concerned general AI topics such as AlphaGo, they were excluded to avoid distortion by temporal heterogeneity. The analytic sample therefore comprised 337 post-ChatGPT records.

Table 1: Construction of the analytic sample

| Step | | Number of threads retained |
|---|---|---|
| 1 | Manually collected unique threads with possible AI-related discussions | 409 |
| 2 | Excluded trivial or irrelevant AI mentions (n=13) | 396 |
| 3 | Excluded records from one repetitive promotional-style user cluster (n=45) | 351 |
| 4 | Excluded records concerning general AI topics before November 2022 (n=14) | 337 |
| | Final analytic sample: post-ChatGPT records | 337 |

With a sample on the order of several hundred cases, this study adopted a mixed-methods approach for subsequent analysis. Descriptive statistics were used to identify overall patterns and subgroups of interest, which were then examined through qualitative close reading of high case-richness threads to reconstruct how those situations unfolded.

### 3.4 Ethics and Reflexivity

The study analyzed publicly accessible discussion threads from an open-access online lymphoma forum. The forum contents were accessible without registration or login, and the study involved no contact with forum members. Under the forum's terms of use, registered users acknowledge that their posts may be viewed by others and used for research purposes, whereas personal information remains private, and the terms permit research use of forum content without further consent [11]. On this basis, no further informed consent was sought. All identifiers were used only for data quality control and were removed from analytic outputs and reporting. Quotations were translated and paraphrased to prevent reverse-identification through search. The study falls within the scope of Article 32 of the Measures for Ethical Review of Life Sciences and Medical Research Involving Humans (National Health Commission, Ministry of Education, Ministry of Science and Technology, and National Administration of Traditional Chinese Medicine, 2023), which exempts research using publicly available, legally obtained, and deidentified data that does not harm individuals or involve sensitive personal information. The author has been an extended observer of the forum, including as the family member of a person with cancer. This positionality afforded familiarity with the community's language, conventions, and authority structures, and it informed both schema development and case selection. This insider position may also shape interpretation; to mitigate this, every reported theme is grounded in documented thread-level evidence.

## 4 RESULTS

### 4.1 How AI Arrived: The Forum's AI Ecology

AI-related discussion was negligible for most of the decade we examined, and then rose steeply with two clear change points that coincide with the public release of ChatGPT in late 2022 and the release of DeepSeek in early 2025 (**Figure 1**). Monthly posting was about 0.14 threads per month before ChatGPT (total 14 threads before November 2022), which increased to roughly 1.6 per month in the ChatGPT era (total 39 threads, December 2022 to December 2024), and to more than 17 per month in the DeepSeek era (total 298 threads after Janurary 2025). Across the three eras, initiators' experiences with AI remained largely favorable, while a small proportion reported unfavorable experiences. Expressions of being overwhelmed were rare before 2025 but became increasingly visible during the DeepSeek era.

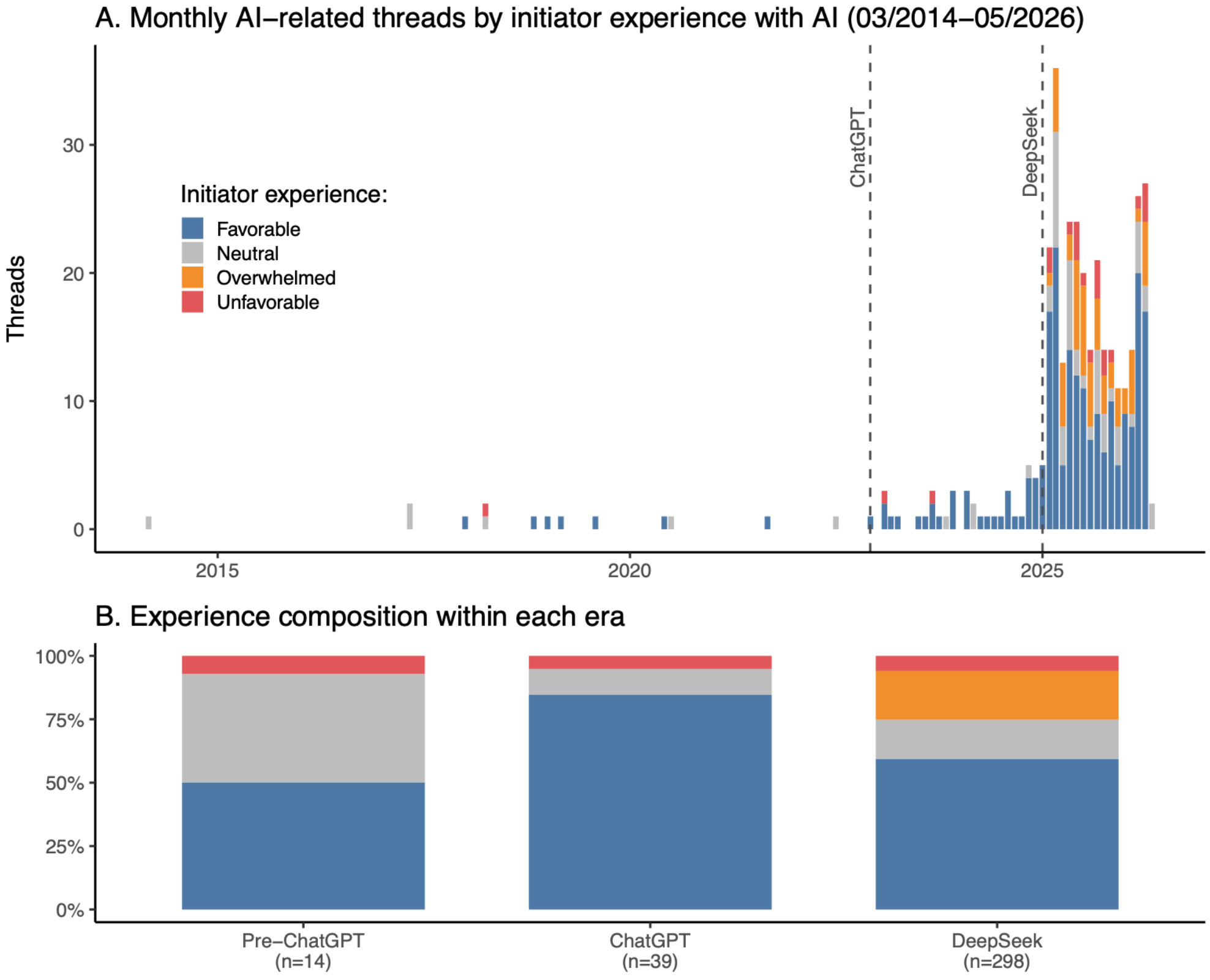

Figure 1: Temporal trends and era composition of AI-related threads, by initiator experience with AI. March 2014 to May 2026. Vertical dashed lines indicate the public release of ChatGPT and DeepSeek.

The meaning of "AI" also shifted across eras. Before ChatGPT, AI mentions were sparse and referred mainly to earlier or more specialized technologies, such as medical-imaging AI. In the ChatGPT era, most discussion concerned conversational tools, especially ChatGPT, while the forum also launched its own AI bot. After the DeepSeek moment, named tools became more diverse, with DeepSeek and Doubao most visible alongside ChatGPT, the forum chatbot, and smaller numbers of other assistants. Across all three eras, however, many members simply wrote "AI" without naming a specific product.

Across the analytic sample (post-ChatGPT eras, n=337), patients introduced AI most often (187, 55.5%), followed by caregivers (118, 35.0%) and community authorities (32, 9.5%). Their experiences with AI were mostly favorable (210, 62.3%), followed by overwhelmed (57, 16.9%), neutral (50, 14.8%), and unfavorable (20, 5.9%) experiences (**Appendix Table A1**). Experience with AI also varied by role. Patients were the most likely to introduce AI favorably, although one in six expressed feeling overwhelmed by AI output. Caregivers were also predominantly favorable, but more than one in five described feeling overwhelmed. Authorities were more often neutral toward AI than patients and caregivers, with none coded as overwhelmed.

Members used AI for far more than looking up medical facts, with many threads combining more than one function (**Appendix Table A2**). Information support was the dominant use (280, 83.1%), followed by second opinion use to check human advice or care decisions (76, 22.6%), and psychosocial support (31, 9.2%). A smaller set of other uses (16, 4.7%) included translation-mediated communication, writing assistance, and broader references to AI as a symbol of future medical progress. Feeling overwhelmed was most common in second-opinion uses (17, 22.4%) and information support uses (54, 19.3%), but was rare in psychosocial support uses (1, 3.2%).

These categories also varied in depth. Information support, for example, ranged from simple report interpretation to more technically involved uses. In one 2024 thread, a patient described deploying a personal AI agent to organize fragmented medical records and track temporal trends in test indicators, suggesting AI-enabled patient agency in illness management.

In addition to genuine AI use, the forum also contained traces of manufactured adoption. One promotional-style account repeatedly recommended two specific AI tools across unrelated threads, contributing over 80% of mentions of both tools in the raw corpus (n=396). The near-identical templated posts were characterized by repetitive boilerplate, stylistic markers atypical of the surrounding community, little relevance to the thread topic, and almost no engagement from other members. This cluster was treated as part of the forum's AI ecology but not as a finding about genuine participant roles, and was therefore excluded from the analytic sample. Nonetheless, its existence showed how easily apparent AI uptake in an OHC can be inflated through low-cost, repetitive, peer-like promotion.

### 4.2 Catching the Patient: Community Uptake of Posting Intent

"Do we still have hope?" In one thread, a young caregiver wrote after their parent was suspected to have lymphoma. The diagnosis was not yet confirmed, but they had asked an AI tool about the radiology report, and received a blunt verdict that survival was unlikely. Frightened, they came to the forum and uploaded the report, hoping other members could check the AI output. What helped, however, was less any single correction than other members saying they had been through the same thing. Community responders shared their fear of the disease and described how they had fought to regain health. Some added that AI had flooded them with frightening readings too, and that this was a known hazard of asking it. Eventually the caregiver was convinced that they should wait for pathology and specialist guidance, and not to be scared off by AI that "often talks nonsense".

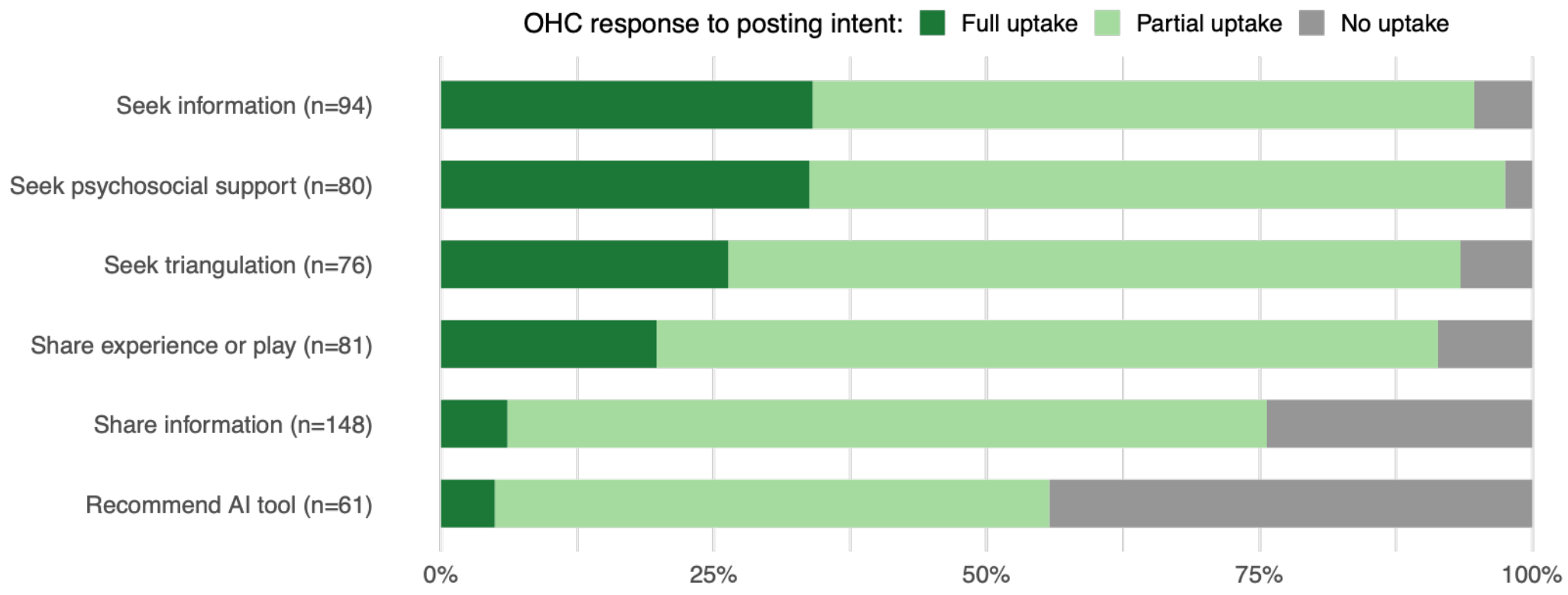

Figure 2: OHC uptake of initiators' posting intent, by posting intent. One thread may involve multiple posting intents.

This was one way in which patients and caregivers brought AI-related content back into the forum for OHC sensemaking. **Figure 2** (**Appendix Table A3**) shows a wider range of initiators' posting intents and how the community took up these intents. Information sharing was the most common reason for AI-related posting (148, 43.9%), followed by seeking information (94, 27.9%), sharing experience or play (81, 24.0%), seeking psychosocial support (80, 23.7%), seeking triangulation (76, 22.6%), and recommending AI tools (61, 18.1%).

Across different posting intents, those seeking psychosocial support, information, or triangulation were almost always answered by the community, with fewer than 7% receiving no uptake. Posts sharing experience or play showed a similar pattern with only 8.6% no uptake. In contrast, posts that shared information or recommended AI tools were easier to pass over, with no-uptake rates of 24.3% and 44.3%, respectively.

Close reading shows why patients and caregivers returned to the community after interacting with AI tools. In some threads, AI multiplied uncertainty rather than resolving it. One caregiver, for example, reported receiving conflicting suggestions from Doubao and DeepSeek about whether to continue a treatment after a troubling side effect. The more answers they compared, the more they had to reconcile. As a result, they had to return to the community for additional evaluation. Members did not substantially adjudicate the disagreement between the tools. Instead, they drew on past experiences with adjustment, monitoring, and when to seek professional advice. The AI dilemma remained only partly addressed, but the caregiver's immediate uncertainty was made more manageable.

A stronger form of uptake appeared when AI and clinical advice pointed in opposite directions. In one thread, a member asked whether lymphoma patients could receive certain treatment for a parallel recurring high-risk condition. Their treating clinician had advised against it, while AI strongly recommended it and supplied references. Experienced members did not simply endorse either side. They distinguished between two treatment forms, noting that one carried specific risks for immunocompromised patients while the other did not, explained why that distinction changed the relevance of the advice, and answered follow-up questions about the member's situation. The member later sought care elsewhere, where clinicians acted in line with the advice that AI and the community had supported. This case showed the forum acting as an informal safety net by making a conflict among authorities more specific, actionable, and clinically usable for its member.

The relatively high uptake of sharing experience or play came from a different mechanism. In one thread, a patient with chemotherapy-related hair loss playfully joked about how AI had accidentally generated a portrait of them with flourishing hair growth. The community replied warmly, wishing for the patient to regain good health, like what AI had depicted.

In another account, a caregiver described how an AI chatbot accompanied their parent through chemotherapy as an emotional confidant, care advisor, and daily companion, helping with diet tracking, protective routines, medication reminders, and late-night conversations that the patient found difficult to have with family members. The caregiver credited AI for healing the enormous emotional trauma induced by cancer, and for supporting the patient tirelessly throughout the treatment journey.

Replies offered appreciation, encouragement, and recognition. Some also expressed interest in trying similar uses themselves.

These cases illustrated how the community could catch crisis and vulnerability. Yet it also had limits in daily availability, especially for private, immediate, and socially awkward forms of micro support. In a thread about everyday uses of AI, one member mentioned uploading embarrassing bodily images to monitor bowel movements, adding that this was not something they felt comfortable sending to a doctor. In such moments, AI was filling a gap the community was structurally unlikely to occupy.

Beyond initiators' intent of posting, their experience with AI also shaped how the community responded. As noted in Section 4.1, around one in six patients and one in five caregivers expressed feeling overwhelmed after interacting with AI. Among these overwhelmed initiators, only 3.5% received no uptake at all and 35.1% received full uptake (**Appendix Table A4**).

## 4.3 But Not the AI: Community Stance Toward the AI Dimension

For the same overwhelmed initiators, however, the community rarely said anything about the AI that had shaped their experience. In fact, across every initiator experience group, the most common community stance toward AI was simply not to address it, ranging from 66.7% of overwhelmed threads to 84.0% of neutral ones (**Appendix Table A5**). Even in threads that presented AI favorably or unfavorably, responders rarely touched the AI topic, leaving it unaddressed in 72.9% and 70.0% of the threads, respectively.

The pattern became clearer when examined against initiators' posting intents (**Figure 3**; **Appendix Table A6**), with unaddressed rates ranging from 62.3% in posts recommending AI tools to 80.9% in posts seeking information. The clearest test came from threads explicitly seeking triangulation, where initiators set competing information sources side by side, including AI, and asked for an adjudication. Even then, the community still tended to bypass AI (71.1% unaddressed) while engaging the medical or emotional substance behind it. A manual review confirmed that in 69 of these 76 threads, AI output itself was among the sources members were asked to judge.

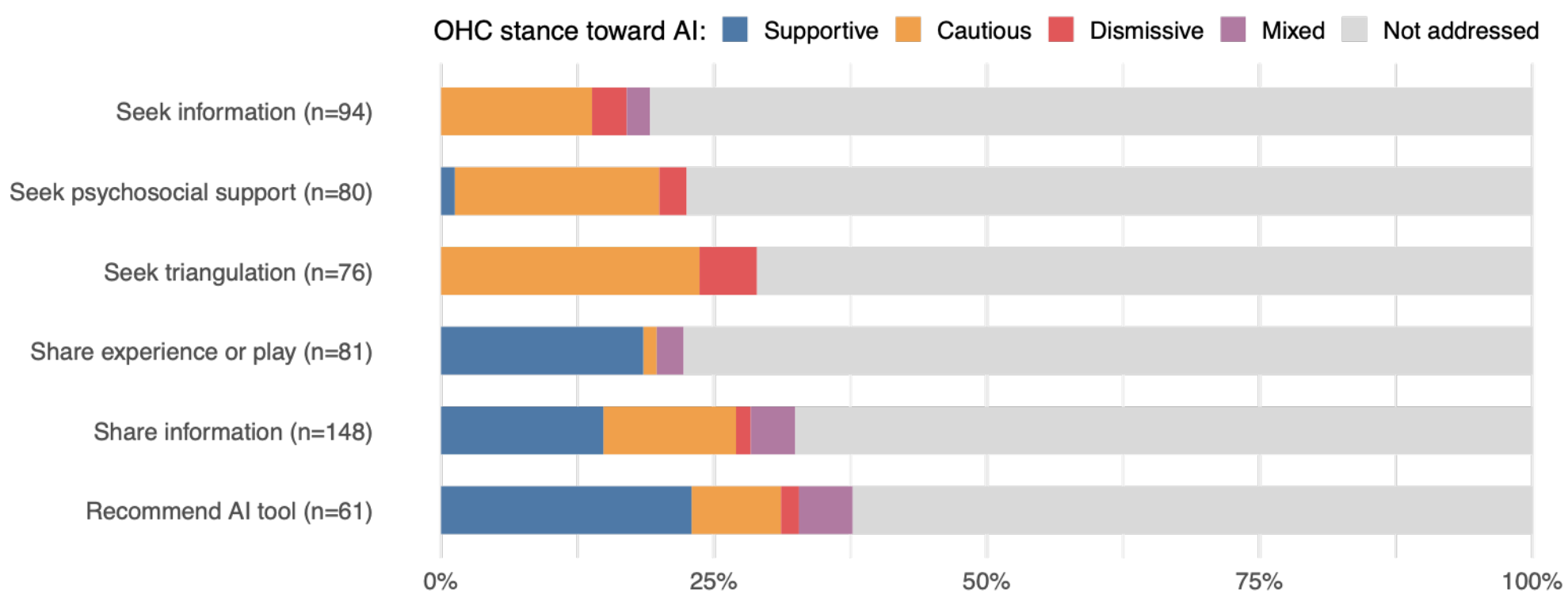

Figure 3: OHC stance toward AI, by posting intent. One thread may involve multiple posting intents.

When the community did engage the AI dimension, the goal was still more to help the initiator settle on a pragmatic agenda, rather than to benchmark AI capabilities. Direct rejection happened rarely, usually when an AI claim carried concrete risk and was externally checkable. For instance, one caregiver asked whether an expensive lymphoma drug had become reimbursable under national policy, later explaining that the information came from an AI chatbot. The community quickly challenged the news, with one member questioning the source after failing to find it online, and another confirming that the drug was still entirely self-paid. A community authority also gave a brief and blunt reply, rejecting the false hope.

Some AI-related risks were easy for the community to flag, as in the reimbursement case above. Others were harder to address, leading to group-level silence in three recurring scenarios: strategic concealment, lay knowledge brokerage, and convenient misattribution.

**Strategic concealment**

Strategic concealment described how members learned to translate backstage AI work before it could safely enter the discussion. In the forum, members regarded self-directed AI use as a way to reduce clinician workload. A member using AI for understanding test results urged others to do the same so they "would not have to ask around." Clinicians, however, tended to read the same behavior as the opposite. In the limited set of forum-published interviews and official posts, clinician voices rarely engaged the question of patient AI use on its own terms. On the few occasions when they did address it directly, their stance was largely unfavorable, treating AI as a source of unfiltered material that complicates rather than supports the consultation. Members responded warmly by thanking them for being caring and dedicated, while leaving the AI topic untouched. This connects directly to the tension in patient-provider communication, and community members were aware of it. In one thread, a member advised others to treat AI assistants as reference opinions to be weighed against the doctor's plan, and warned repeatedly that one should never ever use AI answers to confront their doctor. The warning is an implicit acknowledgment that clinicians may experience being triangulated against AI as a challenge to their authority, and that managing the relationship with the physician is itself part of the work. As a result, a community norm emerged in which members worked around AI but avoided taking a clear stance on whether to endorse or dismiss it.

**Lay knowledge brokerage**

A second scenario occurred when members shifted from consuming AI output to producing and circulating it, presenting themselves as lay knowledge brokers without necessarily providing the information source. Here AI gave members a capacity for synthesis once reserved for clinicians and researchers, but in a form the community's credibility norms may not yet be equipped to vet. Members posted AI-generated summaries of recent drug developments and clinical trials written in an authoritative, review-like register but carrying no traceable citations; promoted particular treatments, including traditional medicine formulas, on the strength of an AI endorsement; and circulated AI-generated hospital or treatment rankings whose omissions other members had to point out in the comments. The common pattern is that some members attach to these posts the surface form of AI-generated expert knowledge while detaching them from the sources that would let anyone check them. In response, the community rarely allocated its attention to lengthy mechanical content that was visibly AI-generated. When it did engage, the focus was usually to address the disease-related claim at stake rather than the role of AI.

**Convenient misattribution**

This reluctance to evaluate AI as the main object had an uneven consequence. When AI seemed misleading or wrong, the community had little reason to escalate the alarm. When AI provided useful information, however, its contribution could be buried under other forms of judgment. In the AI-versus-doctor case of Section 4.2, AI supplied references for a treatment that differed from the treating clinician's advice. The community discussion and later clinical endorsement both helped the member act in line with the AI-supported recommendations, yet no participants treated AI as a source to be credited.

This pattern could also work through blame. Recall the opening vignette, where a member was frightened by an AI interpretation of a dangerously low laboratory value. One respondent reassured them by suggesting that AI had misread the report. Yet on later inspection of the uploaded image, the laboratory value itself was abnormal enough to warrant concern, and no apparent OCR error was found through manual inspection. That reassurance may have helped redirect panic toward action, but it also placed the error in AI without checking whether AI had actually made an error. Other responses, while actively providing support, did not verify whether the AI reading was right or wrong. These cases illustrated convenient misattribution, where the community took AI as something that shaped a thread, but not as important enough to receive careful credit or careful blame.

The community's tendency to leave AI unaddressed also extended to its authority layer. Forum administrators were less often favorable and more neutral than other roles when they initiated AI-related topics (**Appendix Table A1**), but their positions evolved over time and shifted by situation, sometimes using AI while also warning members against it.

This unsettled stance was visible across several long-tenured authority accounts. Early uses of AI in 2024 tended to be light and low-stakes, for example treating AI image generation as community entertainment. AI use was also disclosed in some responses to members' medical queries. In the same year, the forum launched its own AI chatbot and promoted it as an assistant for interpreting medical reports. However, the bot never became a sustained part of community practice. Within a few months, references to it quietly tapered off, without the administrators pushing it further or members asking for its return. In early 2025, some authority figures quickly adopted DeepSeek after its release to support members' information needs. As DeepSeek-era use intensified, however, they also cautioned members against turning to AI in moments of prognostic anxiety, arguing that it could produce errors and amplify distress. At the same time, they continued to treat lower-stakes uses such as writing and translation as helpful, and posted informal guidance on when and how the AI tools may best serve members. By 2026, some authorities again directed newly diagnosed members to ask AI when forum responses were limited, while others appeared to assume that members had likely already consulted AI before posting, and adjusted their replies accordingly. Across these cases, authority engagement with AI remained unfinished and situational.

Across the analytic sample (n=337), the community took up initiators' posting intents at least partially, in 81.6% of the threads, while leaving the AI dimension untouched in 73.3% (**Table 2**). Among threads with partial uptake (n=211), AI was not addressed in 70.1%. Even among threads with full uptake (n=64), AI was still not addressed in 62.5%. This contrast captures the central pattern across the results: the forum reliably caught the patient, but not the AI.

Table 2: OHC stance toward AI, by OHC uptake of initiator's posting intent

| | AI not addressed | OHC other stances | Row Sums |
|---|---|---|---|
| OHC no uptake | 59 (95.2%) | 3 (4.8%) | 62 |
| OHC partial uptake | 148 (70.1%) | 63 (29.9%) | 211 |
| OHC full uptake | 40 (62.5%) | 24 (37.5%) | 64 |

[a] Cells are subgroup numbers (row %). OHC other stances include: supportive, cautious, dismissive, and mixed.

# 5 DISCUSSION

## 5.1 Catch the Patient, Not the AI

Taken together, the findings show that the decisive interpretive work increasingly happens after AI has spoken, in the social space where members decide what an AI answer means for them. In some threads members paste or reference AI output and ask peers or administrators to verify it. In others, AI has already shaped a member's anxiety or interpretation before the discussion begins, and the community responds to the medical concern rather than to AI as such. Either way, AI did not replace community work, but moved the locus of sensemaking downstream.

The shift was possible because AI entered the treatment journey through functions that OHCs had long performed [1-4, 13, 14]. The clearest evidence is the overlap between what members used AI for and why they had traditionally posted. Information support, the single most common use of AI in our sample, is also one of the oldest reasons people come to an OHC at all [13, 14]. When a member uploads a pathology report to a chatbot rather than asking the forum to explain it, AI is absorbing a function that peers used to perform. In this narrow sense AI does displace some community work. The same point also reframes AI's role as a psychosocial support. Members used AI as a confidant, as an emotional companion, as a way of producing images and language that expressed their situation, and as a vehicle for hope about the future. These uses are about sensemaking of a life with illness, and they are largely absent from medical-AI work centered on diagnosis and treatment.

What our data show, however, is that the displacement is not always the end of the story. Often AI produced more information, more possibilities, and sometimes conflicting advice, generating fresh cognitive work of filtering, comparison, and integration. In several cases AI created new decision work rather than dissolving the original burden. As a result, members who used AI for information might carry the AI's output back into the forum to do something AI had not settled for them, most often to triangulate it against human judgment or to seek reassurance about what it meant. The traditional

OHC function did not disappear; it moved one step downstream, from producing an interpretation to adjudicating one that AI had already produced.

That reframing changes which question matters for CSCW. Model accuracy still matters, but the issue raised in our setting is that accuracy alone does not settle the work created by AI. Once AI produces a candidate interpretation, someone still has to place it in context, calibrate it against lived experience and clinical authority, and make it socially usable. In this forum that work fell, when it fell to anyone, on the community. But it did so unevenly and informally, depending on whether experienced members were present and whether anyone took up the AI dimension at all.

### 5.2 Community Norms for AI-Related Participation

As AI moves sensemaking downstream, the next question is whether existing support practices and credibility infrastructure are sufficient for AI-related work. OHC norms are not directly transferable across platforms or health conditions. For example, responsible participation in some communities requires avoiding prescriptive guidance [18]. In this lymphoma forum, however, discussion of treatment regimens, pathology interpretation, and adverse effects was one of the community's core support functions. A general norm against "medical advice" would therefore miss what members actually need from one another. Still, other norms, such as citing sources and sharing lived experience, remain important across different settings. The question, then, is how AI changes the basis on which peer contributions become credible.

Existing OHC credibility practices evolved around people rather than AI. Members learn to judge contributions by who is speaking, whether claims are grounded in lived experience, and how they are taken up or corrected by experienced peers [1, 3]. AI-generated content does not fit neatly into this credibility infrastructure. It has no lived experience to share, nor does it claim to do so. Such content therefore often falls outside existing credibility checks, leaving the AI dimension unaddressed rather than explicitly accepted or rejected by the community.

Lay knowledge brokerage makes this problem especially visible. AI allowed members to produce polished summaries without necessarily providing traceable sources or disclosing where the synthesis came from. These posts were not always malicious, and some may have been intended as help. Yet they entered the forum with the surface form of expertise while bypassing ordinary credibility cues. The promotional cluster excluded during data cleaning represents a crude version of the same problem, where AI-flavored, templated participation was manufactured to mimic genuine community participation and peer recommendation. As AI-written contributions become harder to distinguish from lived-experience posts, a community that runs on experiential credibility faces a signal it is not yet equipped to read, and the burden of detection currently falls on whichever experienced member happens to notice [17].

Convenient misattribution is a second, subtler failure. Even when members do notice AI's role, the forum has no practice of assigning it careful credit or blame. In many cases, AI was treated as background when it helped, or as a convenient explanation when it amplified confusion, without becoming an object of careful evaluation in either direction. The result was that AI-related practices stayed fragmented across individual threads, instead of accumulating into community knowledge useful even for members who did not join the thread.

AI changed what must be made visible for peer support to remain credible. OHCs thus need new norms for AI-related participation, for example by asking where a synthesis comes from, whether it can be checked, and what role AI played in producing it. Such norms would complement lived experience with provenance, helping members recognize when AI should be questioned or corrected.

### 5.3 Agency is Collectively Negotiated

Recent work has shown how LLMs reshape patient agency throughout individual healthcare-seeking journeys [6]. The present study extends this perspective by examining what happens after AI-mediated agency enters an existing community.

Existing accounts of collective sensemaking would suggest that an object as ambiguous, novel and consequential as AI should itself invite lateral engagement and collective interpretation, similar to how controversial treatments and conflicting information were handled in the case of TuDiabetes [4]. The forum did otherwise. Although community sensemaking remained actively organized around the patient's situation, AI rarely became an object of collective negotiation. In 73.3% of the threads the community stance was simply not to address AI. Even when members explicitly sought triangulation, the not-addressed rate stayed above 70%.

This provides a boundary condition, rather than a counterexample, to the existing framework on collective sensemaking. The framework explains how communities collectively interpret uncertainty, but leaves open what gets selected as the object of that interpretation. Our findings suggest that this selection is fundamentally person-centered. The community mobilizes around people in need, but not the tools they bring with them. AI therefore occupies a third position, neither an object of sensemaking, nor an irrelevant bystander, but an upstream force that had already shaped members' questions, anxieties, and framings before they arrived at the forum. By doing so, AI shifted the locus of collective sensemaking downstream, from interpreting information itself to interpreting people whose situations AI had shaped. This is the locus shift in full: the community continues to catch the person even when it does not catch the AI.

What follows is an interpretive authority that settles in no single place. OHC members rarely organized this social re-entry as a contest between the doctor and the AI, but more often as an orchestration of multiple sources. A member would hold a clinician's recommendation, an AI explanation, peer experience, and sometimes the published literature side by side and try to compose a coherent picture from them. This is where agency became collective. A design question follows: how to support the wider assembly of human and machine sources that a patient coordinates, and the labor of holding that assembly together.

Yet this collective agency was often accompanied by strategic concealment. Patients sometimes used AI to reduce the burden placed on physicians, but also warned one another not to let the physician know they had done so. This may point to a coordination failure around a tool that both parties increasingly know is in use, but that has not yet found a legitimate place in patient-clinician communication. Patients therefore had to manage two relationships at once, one with the clinician and one with the AI, hiding the second to protect the first. Viewed from another angle, the problem also points to an opportunity. If patient AI use were treated as a shared and expected input rather than a transgression to be hidden, the same activity could become a basis for collaboration, with the community, and potentially the clinician, helping the patient use AI well rather than leaving them to do it in secret.

### 5.4 Implications

The findings suggest that patient-facing AI should be designed on the assumption that its output will travel, and that context may be lost in transit. Members are going to use AI to make sense of their illness whether or not designers intend it, so the useful question is not how to discourage this but how to make it safer when it happens.

One direction is to support OHCs in building AI literacy. The risks surfaced from the forum data, including lay knowledge brokerage, convenient misattribution, and promotion disguised as participation, are all failures that arise when AI output enters a community not prepared to handle it. Establishing OHC norms around disclosure, source checking, promotional boundaries, and proper recognition of AI's role gives both the topic initiator and respondents a fairer chance of catching its errors [17, 18].

A second direction is AI for multi-party care work, rather than as a private chatbot or a competitor to clinicians. Strategic concealment often occurs when legitimate uses get pushed backstage, leaving patients to manage AI alone. A safer response may be to design AI to complement support patients already assemble around themselves, especially where formal care, peer support, and family support are unevenly available.

The findings also point to the need for designs that are aware of users' emotional states. Harm did not only come from wrong answers. Medically plausible information could also intensify distress if delivered as a blunt prognosis, while generic and heavily hedged answers could leave overwhelmed users with too much uncertainty and no clear next step. Patient-facing AI should therefore be sensitive to the state of the person receiving the answer, both informationally and emotionally, so that its answers become usable rather than merely available.

### 5.5 Limitations

Several limitations bound our claims. First, although the author manually verified every record against the original threads, the structured coding relied on a single coder working with LLM-assisted extraction, and a formal second-coder reliability audit remains future work. Second, the design is largely cross-sectional and observational, which supports description of patterns and associations but not causal claims about how AI affects members' distress or decisions. Third, we observe only what members chose to bring back to the forum, not their full interactions with AI, their private deliberations, or their clinical outcomes. We therefore cannot characterize the everyday AI experience of lymphoma patients and caregivers in general. We can say only that the members who posted reported particular uses and experiences, while those who never posted, or whose AI use never surfaced in the forum, are absent from our view. Finally, the study draws on a single Chinese, disease-specific community, which might shape how AI tools and dynamics appear. Transferring these findings to other diseases, platforms, and health systems calls for care.

## 6 CONCLUSION

Patient-facing AI should not be evaluated only at the point of individual use. Its consequences continue to unfold afterward, in the peer communities where people exchange knowledge and experiences to support one another. In this forum, AI became neither a substitute for collective sensemaking nor a stable object of community scrutiny. Instead, it changed where that interpretive work happened. Members carried AI-related content back into the forum, and other members engaged the person and their illness far more readily than the AI itself. In doing so, the community provided a space where members collectively made sense of illness, managed life with cancer, and recovered from moments when an AI encounter left them frightened or adrift. For CSCW, the implication is that OHCs remain important sites of illness sensemaking even when AI enters the pathway. The opportunity lies in strengthening that collective capacity, supporting communities as sites where people help one another become competent and resilient users of AI.

## ACKNOWLEDGMENTS

The author would like to thank members of the online patient forum for sharing their experiences, which made this research possible. ChatGPT-5.5 and Claude Opus 4.8 were used for language editing, and to assist with code development. All scientific content and analyses were conducted and verified by the author.

## REFERENCES

[1] A. Hartzler and W. Pratt, "Managing the personal side of health: how patient expertise differs from the expertise of clinicians," Journal of medical Internet research, vol. 13, no. 3, p. e62, 2011.

[2] J. Huh, R. Patel, and W. Pratt, "Tackling dilemmas in supporting 'the whole person' in online patient communities," in Proceedings of the SIGCHI

Conference on Human Factors in Computing Systems, 2012, pp. 923-926.

[3] J. Huh and W. Pratt, "Weaving clinical expertise in online health communities," in Proceedings of the SIGCHI conference on human factors in computing systems, 2014, pp. 1355-1364.

[4] L. Mamykina, D. Nakikj, and N. Elhadad, "Collective sensemaking in online health forums," in Proceedings of the 33rd Annual ACM Conference on Human Factors in Computing Systems, 2015, pp. 3217-3226.

[5] A. Montero, J. M. III, A. Kearney, I. Valdes, A. Kirzinger, and L. Hamel. "KFF Tracking Poll on Health Information and Trust: Use of AI For Health Information and Advice." https://www.kff.org/public-opinion/kff-tracking-poll-on-health-information-and-trust-use-of-ai-for-health-information-and-advice/ (accessed 29 Jun, 2026).

[6] Y. Cao et al., "More than Decision Support: Exploring Patients' Longitudinal Usage of Large Language Models in Real-World Healthcare-Seeking Journeys," in Proceedings of the 2026 CHI Conference on Human Factors in Computing Systems, 2026, pp. 1-24.

[7] J. W. Ayers et al., "Comparing physician and artificial intelligence chatbot responses to patient questions posted to a public social media forum," JAMA internal medicine, vol. 183, no. 6, pp. 589-596, 2023.

[8] S. Shekar, P. Pataranutaporn, C. Sarabu, G. A. Cecchi, and P. Maes, "People overtrust AI-generated medical advice despite low accuracy," NEJM AI, vol. 2, no. 6, p. AIoa2300015, 2025.

[9] A. Hassoon et al., "Evaluating the AI Potential as a Safety Net for Diagnosis: A Novel Benchmark of Large Language Models in Correcting Diagnostic Errors," medRxiv, p. 2026.02. 22.26346832, 2026.

[10] M. Khosravi, Z. Zare, S. M. Mojtabaeian, and R. Izadi, "Artificial intelligence and decision-making in healthcare: a thematic analysis of a systematic review of reviews," Health services research and managerial epidemiology, vol. 11, p. 23333928241234863, 2024.

[11] K. Ning et al., "Online Health-Seeking Behaviors and Information Needs Among Patients With Lymphoma in China: Study of Regional and Temporal Trends," Journal of Medical Internet Research, vol. 27, p. e80497, 2025.

[12] E. Dong, J. Xu, X. Sun, T. Xu, L. Zhang, and T. Wang, "Differences in regional distribution and inequality in health-resource allocation on institutions, beds, and workforce: a longitudinal study in China," Archives of Public Health, vol. 79, no. 1, p. 78, 2021.

[13] S. Ziebland and S. Wyke, "Health and illness in a connected world: how might sharing experiences on the internet affect people's health?," The Milbank Quarterly, vol. 90, no. 2, pp. 219-249, 2012.

[14] Y.-C. Wang, R. E. Kraut, and J. M. Levine, "Eliciting and receiving online support: using computer-aided content analysis to examine the dynamics of online social support," Journal of medical Internet research, vol. 17, no. 4, p. e99, 2015.

[15] L. Salmi et al., "A proof-of-concept study for patient use of open notes with large language models," JAMIA open, vol. 8, no. 2, p. ooaf021, 2025.

[16] V. Sorin et al., "Large language models and empathy: systematic review," Journal of medical Internet research, vol. 26, p. e52597, 2024.

[17] Y. Shulman, A. Kitkowska, and M. Warner, "Discerning Authorship in Online Health Communities: Experience, Trust, and Transparency Implications for Moderating AI," arXiv preprint arXiv:2604.19429, 2026.

[18] S. Mittal et al., "Follow the Rules (or Not): Community Norms and AI-Generated Support in Online Health Communities," arXiv preprint arXiv:2603.19093, 2026.

[19] L. Wang et al., "Cass: Towards building a social-support chatbot for online health community," Proceedings of the ACM on Human-Computer Interaction, vol. 5, no. CSCW1, pp. 1-31, 2021.

# A APPENDICES

## A.1 Methodological Details

### *A.1.1 Search Terms Used for Data Collection*

The following search terms were used to identify AI-related discussion threads on House086: DeepSeek, Doubao (“豆包”), Qwen (“千问”), Kimi, Ant A-Fu (“蚂蚁阿福”), iFLYTEK (“讯飞”), Xiaohe Doctor (“小荷医生”), ERNIE Bot (“文心一言”), ChatGPT (“GPT”), Yuanbao (“元宝”), AI, artificial intelligence (“人工智能”), Claude, Gemini, MiniMax, large language model (“大模型”), and machine learning (“机器学习”).

### *A.1.2 Operational Definitions Used in LLM-Assisted Extraction*

1. AI use case
   Information support = Using AI to explain, search, organize, or navigate medical or care-related information. Second opinion = Using AI to check, compare, challenge, or support medical, hospital, or treatment decisions. Psychosocial

support = Using AI for comfort, companionship, encouragement, or playful/creative support. Other = Any other practical use.

2. Initiator = The first user in a House086 discussion thread who explicitly mentioned AI, regardless of whether the mention occurred in the original post or in a reply.
3. Authority = Senior community members, including forum administrators, experienced patient advocates, the forum bot, and physicians posting through official forum accounts.
4. Initiator experience with AI

   Favorable = The initiator expressed support for, trust in, or reliance on AI output; considered AI helpful or satisfactory; recommended AI tools to others; or used AI conclusions to inform decision-making.

   Overwhelmed = AI output triggered or intensified emotional breakdown, anxiety, panic, information overload, feeling overwhelmed, or decision paralysis.

   Unfavorable = The initiator opposed, criticized, or rejected AI output; reported being harmed or misled by AI; or advised others not to trust AI.

   Neutral = The initiator used AI without expressing any evaluation, recommendation, or clear emotional attitude toward it.

   Note: Coding should not rely solely on the tone used to describe AI output in the OHC. Instead, it should reflect the actual impact of the AI output on the initiator's subsequent behavior and psychological state.
5. Initiator post intent

   Seek information = Request general medical or care-related information from the community.

   Seek triangulation = Request the community to judge between AI, physicians, or other information sources (e.g., conflicting advice from AI and a physician).

   Seek psychosocial support = Seek reassurance, comfort, or emotional support.

   Share information = Share information provided by AI or opinions about AI (knowledge sharing).

   Share experience or play = Share personal experiences, creative uses, hope, or companionship involving AI.

   Recommend AI tool = Recommend an AI tool to other community members. Other = Any other practical posting intent.
6. OHC uptake of the initiator's post intent

   Yes = The community clearly addressed the initiator's primary posting intent. For help-seeking posts, the community provided substantive assistance. For sharing or recommendation posts, the community engaged meaningfully with the content.

   Partial = The community responded to part of the initiator's intent or to part of the thread, but did not fully address the primary intent, or only engaged at a superficial level.

   No = Subsequent replies did not address the initiator's primary intent, or no replies were posted.

   Note: Uptake was coded as a third-party judgment of whether replies addressed the stated intent, without inferring community capacity or initiator satisfaction.
7. OHC stance toward AI

   Supportive = Replies validated, endorsed, or built upon the use of AI.

   Cautious = Replies hedged, qualified, advised restraint, expressed skepticism, suggested verification, or redirected the initiator to authoritative sources.

   Dismissive = Replies explicitly rejected or disparaged AI (e.g., "AI is nonsense" or "Don't ask AI").

   Mixed = Different replies expressed different stances toward AI.

Not addressed = No reply addressed the AI dimension, either because there were no replies or because replies focused on other issues while ignoring the AI mention.

Note: OHC uptake of the initiator's post intent and OHC stance toward AI were coded independently. For example, the community may respond to the patient's medical concern (uptake = yes) while not addressing AI at all (stance = not addressed).

8. AI case richness (1–5): AI case richness was used to identify cases for qualitative close reading. The levels were cumulative, meaning that a higher level required satisfying all lower levels.

   1 = Only an AI tool name or a generic reference to AI was mentioned, with no specific use case.

   2 = An AI use case involving the initiator or another person was mentioned, but without substantive detail.

   3 = The thread contained concrete AI-related content, such as AI outputs, usage experiences, emotional reactions, or clear attitudes.

   4 = In addition to Level 3, the thread showed evidence that AI shaped how the initiator participated in the OHC.

   5 = In addition to Level 4, other OHC members substantively engaged with the AI-related content by discussing, verifying, questioning, supplementing, or rejecting it.

### A.2 Supplementary Tables

Table A1: Initiator experience with AI, by initiator identity

| | Favorable | Neutral | Overwhelmed | Unfavorable | Row Sums |
|---|---|---|---|---|---|
| Patient | 132 (70.6%) | 17 (9.1%) | 31 (16.6%) | 7 (3.7%) | 187 |
| Caregiver | 65 (55.1%) | 18 (15.3%) | 26 (22.0%) | 9 (7.6%) | 118 |
| Authority | 13 (40.6%) | 15 (46.9%) | 0 (0.0%) | 4 (12.5%) | 32 |
| Column Sums | 210 (62.3%) | 50 (14.8%) | 57 (16.9%) | 20 (5.9%) | 337 |

[a] Cells are subgroup numbers (row %). Of the 337 threads, 9.5% were initiated by authority, 35% by caregivers, and 55.5% by patients.

Table A2: Initiator experience with AI, by AI use category

| | Favorable | Neutral | Overwhelmed | Unfavorable | Row Sums |
|---|---|---|---|---|---|
| Information support | 174 (62.1%) | 36 (12.9%) | 54 (19.3%) | 16 (5.7%) | 280 |
| Second opinion | 42 (55.3%) | 10 (13.2%) | 17 (22.4%) | 7 (9.2%) | 76 |
| Psychosocial support | 26 (83.9%) | 4 (12.9%) | 1 (3.2%) | 0 (0.0%) | 31 |
| Other | 8 (50.0%) | 6 (37.5%) | 1 (6.2%) | 1 (6.2%) | 16 |

[a] Cells are subgroup numbers (row %). One thread may involve multiple AI use categories. Total 337 threads.

Table A3: OHC uptake of initiator's posting intent, by initiator's intent

| | OHC no uptake | OHC partial uptake | OHC full uptake | Row Sums |
|---|---|---|---|---|
| Seek information | 5 (5.3%) | 57 (60.6%) | 32 (34.0%) | 94 |
| Seek psychosocial support | 2 (2.5%) | 51 (63.8%) | 27 (33.8%) | 80 |
| Seek triangulation | 5 (6.6%) | 51 (67.1%) | 20 (26.3%) | 76 |
| Share information | 36 (24.3%) | 103 (69.6%) | 9 (6.1%) | 148 |
| Share experience or play | 7 (8.6%) | 58 (71.6%) | 16 (19.8%) | 81 |
| Recommend AI tool | 27 (44.3%) | 31 (50.8%) | 3 (4.9%) | 61 |

[a] Cells are subgroup numbers (row %). One thread may involve multiple posting intents. Total 337 threads.

Table A4: OHC uptake of initiator's intent by initiator's experience with AI

| | OHC no uptake | OHC partial uptake | OHC full uptake | Row Sums |
|---|---|---|---|---|
| Favorable | 52 (24.8%) | 128 (61.0%) | 30 (14.3%) | 210 |
| Neutral | 5 (10.0%) | 37 (74.0%) | 8 (16.0%) | 50 |
| Overwhelmed | 2 (3.5%) | 35 (61.4%) | 20 (35.1%) | 57 |
| Unfavorable | 3 (15.0%) | 11 (55.0%) | 6 (30.0%) | 20 |
| Column Sums | 62 (18.4%) | 211 (62.6%) | 64 (19.0%) | 337 |

[a] Cells are subgroup numbers (row %).

Table A5: OHC stance toward AI, by initiator's experience with AI

| | Not addressed | OHC supportive | OHC cautious | OHC dismissive | Mixed stances | Row Sums |
|---|---|---|---|---|---|---|
| Favorable | 153 (72.9%) | 29 (13.8%) | 20 (9.5%) | 3 (1.4%) | 5 (2.4%) | 210 |
| Neutral | 42 (84.0%) | 3 (6.0%) | 5 (10.0%) | 0 (0.0%) | 0 (0.0%) | 50 |
| Overwhelmed | 38 (66.7%) | 0 (0.0%) | 16 (28.1%) | 3 (5.3%) | 0 (0.0%) | 57 |
| Unfavorable | 14 (70.0%) | 0 (0.0%) | 3 (15.0%) | 2 (10.0%) | 1 (5.0%) | 20 |
| Column Sums | 247 (73.3%) | 32 (9.5%) | 44 (13.1%) | 8 (2.4%) | 6 (1.8%) | 337 |

[a] Cells are subgroup numbers (row %).

Table A6: OHC stance toward AI, by initiator's posting intent

| | Not addressed | OHC supportive | OHC cautious | OHC dismissive | Mixed stances | Row Sums |
|---|---|---|---|---|---|---|
| Seek information | 76 (80.9%) | 0 (0.0%) | 13 (13.8%) | 3 (3.2%) | 2 (2.1%) | 94 |
| Seek psychosocial support | 62 (77.5%) | 1 (1.2%) | 15 (18.8%) | 2 (2.5%) | 0 (0.0%) | 80 |
| Seek triangulation | 54 (71.1%) | 0 (0.0%) | 18 (23.7%) | 4 (5.3%) | 0 (0.0%) | 76 |
| Share information | 100 (67.6%) | 22 (14.9%) | 18 (12.2%) | 2 (1.4%) | 6 (4.1%) | 148 |
| Share experience or play | 63 (77.8%) | 15 (18.5%) | 1 (1.2%) | 0 (0.0%) | 2 (2.5%) | 81 |
| Recommend AI tool | 38 (62.3%) | 14 (23.0%) | 5 (8.2%) | 1 (1.6%) | 3 (4.9%) | 61 |

[a] Cells are subgroup numbers (row %). One thread may involve multiple posting intents. Total 337 threads.